\documentclass[a4paper,fleqn,usenatbib]{mnras}

\usepackage{mathptmx}
\usepackage{txfonts}

\usepackage[T1]{fontenc}
\usepackage{ae,aecompl}
\usepackage{multicol}
\usepackage{graphicx}
\usepackage{color}
\usepackage{graphicx}
\usepackage{amssymb}
\usepackage{rotating}
\usepackage{longtable}
\usepackage{lscape}

\newcommand{\figref}{Figure~\ref}

\title[NGC\,7492 and the Sagittarius tidal stream]{The globular cluster NGC\,7492 and the Sagittarius tidal stream: together but unmixed}

\author[J. A. Carballo-Bello et al.]{
J. A. Carballo-Bello$^{1,2}$\thanks{E-mail: jcarballo@astro.puc.cl}, 
J. M. Corral-Santana$^{3}$, M. Catelan$^{1,2}$, D. Mart\'inez-Delgado$^{4}$,\newauthor 
R. R. Mu\~noz$^{5}$, A. Sollima$^{6}$, C. Navarrete$^{1,2}$, S. Duffau$^{7}$, P. C\^ot\'e$^{8}$ \& M. D. Mora$^{1}$ \\
\\
$^{1}$Instituto de Astrof\'isica, Facultad de F\'isica, Pontificia Universidad Cat\'olica de Chile, Av. Vicu\~na Mackenna, 4860, 782-0436,\\
Macul, Santiago, Chile\\
$^{2}$Millenium Institute of Astrophysics, Santiago, Chile\\
$^{3}$European Southern Observatory, Alonso de C\'ordova 3107, Vitacura, Santiago, Chile\\
$^{4}$Astronomisches Rechen-Institut, Zentrum f\"ur Astronomie der Universit\"at Heidelberg, M\"onchhofstr. 12-14, D-69120 Heidelberg, Germany\\
$^{5}$Departamento de Astronom\'ia, Universidad de Chile, Camino El Observatorio 1515, Las Condes, Santiago, Chile\\
$^{6}$INAF Osservatorio Astronomico di Bologna, via Ranzani 1, I-40127 Bologna, Italy\\
$^{7}$Departamento de Ciencias F\'isicas, Universidad Andr\'es Bello, Fern\'andez Concha 700, Las Condes, Santiago, Chile\\
$^{8}$National Research Council of Canada, Herzberg Astronomy and Astrophysics, Victoria, BC, V9E 2E7, Canada\\
}

\date{Accepted XXX. Received YYY; in original form ZZZ}

\pubyear{2017}

\begin{document}
\label{firstpage}
\pagerange{\pageref{firstpage}--\pageref{lastpage}}
\maketitle

\begin{abstract}

We have derived from VIMOS spectroscopy the radial velocities for a sample of 71 stars selected from CFHT/Megacam photometry around the Galactic globular cluster NGC\,7492. In the resulting velocity distribution, it is possible to distinguish two relevant non-Galactic kinematic components along the same line of sight: a group of stars at $\langle{v_{\rm r}}\rangle \sim 125$\,km\,s$^{-1}$ which is compatible with the velocity of the old leading arm of the Sagittarius tidal stream, and a larger number of objects at $\langle{v_{\rm r}}\rangle \sim -110$\,km\,s$^{-1}$ that might be identified as members of the trailing wrap of the same stream. The systemic velocity of NGC\,7492 set at $v_{\rm r} \sim -177$\,km\,s$^{-1}$ differs significantly from that of both components, thus our results confirm that this cluster is not one of the globular clusters deposited by the Sagittarius dwarf spheroidal in the Galactic halo, even if it is immersed in the stream. A group of stars with $<v_{\rm r}> \sim -180$\,km\,s$^{-1}$ might be comprised of cluster members along one of the tidal tails of NGC\,7492.

\end{abstract}

\section{Introduction}

The accretion of the Sagittarius (Sgr) dwarf galaxy \citep{Ibata1994} and the stellar tidal stream that its disruption has generated around the Milky Way \citep[e.g.][]{Martinez-Delgado2001,Newberg2002,Majewski2003,Belokurov2006a,Koposov2012,Huxor2015,Navarrete2017a}  represent one of the best examples of ongoing accretion of satellite galaxies in the Local Universe. Such events contribute to the host galaxy halo not only with stars but also with globular clusters \citep[GCs; e.g.][]{Leaman2013,Zaritsky2016}.

Indeed, substantial evidence of the accretion of GCs in the Milky Way have been gathered in recent years. In the case of Sgr, at least 4 globulars are found in its main body \citep[M\,54, Arp\,2, Terzan\,7 and Terzan\,8;][]{DaCosta1995} and a few halo GCs have been associated with the stream across the sky \citep[e.g.][]{Dinescu2000,Bellazzini2002,Palma2002, Martinez-Delgado2002,Bellazzini2003,Carraro2009,Forbes2010,Dotter2011,Sbordone2015}. \cite{Bellazzini2003} estimated that $\sim 20\%$ of halo GCs beyond $R_{\rm G} = 10$\,kpc might be clusters formed in the interior of Sgr and later accreted by our Galaxy, while \citet[][hereafter LM10a]{Law2010b} proposed a list of 9 of these systems spatially and kinematically compatible with the predicted path of the stream \citep[][hereafter LM10b]{Law2010a}.

 \begin{figure*}
     \begin{center}
      \includegraphics[scale=0.38]{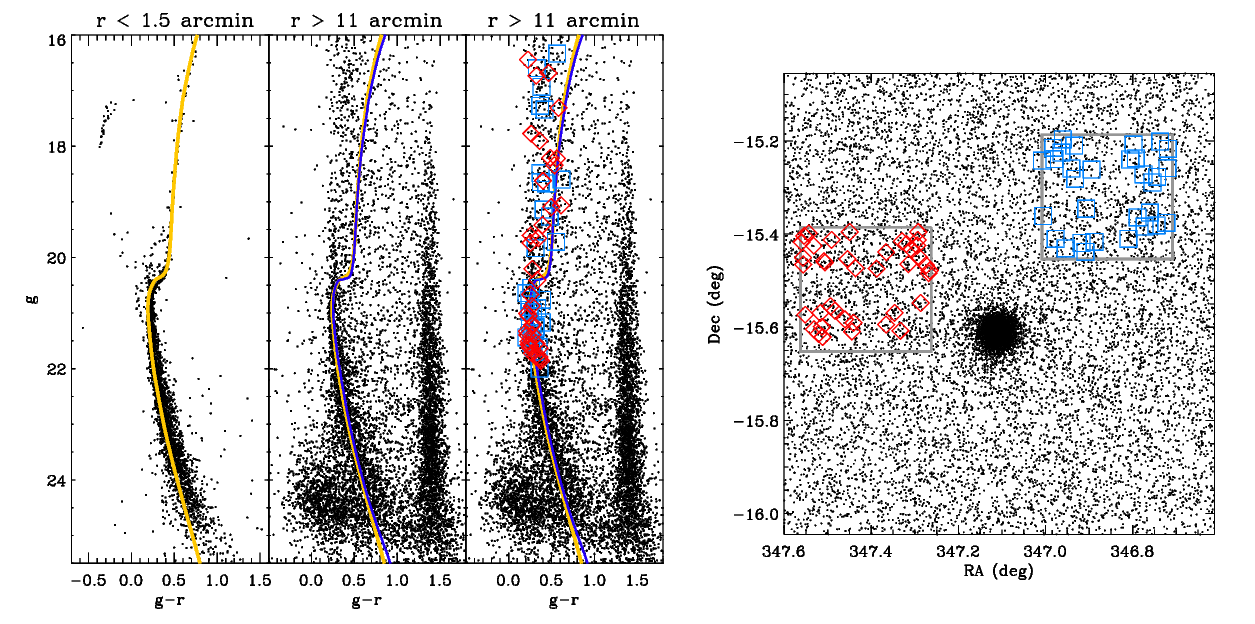}
      \caption[]{\emph{First and second panels from the left:} Megacam CMDs corresponding to the stars in the catalog at distances $r < 1.5$\,arcmin and $r > 11$\,arcmin from the center of NGC\,7492, respectively. The isochrone fitting shows that the underlying main-sequence observed in the second panel seems to be associated with a stellar system at $d_{\odot} \sim 26$\,kpc. The yellow solid line corresponds to a $t \sim 12$\,Gyr and [Fe/H] $\sim -1.8$ population while the blue one corresponds to $t \sim 10$\,Gyr and [Fe/H] $\sim -1.5$. \emph{Third panel from the left:} same CMD for stars with $r > 11$\,arcmin where the target stars in Fields 1 and 2 are overplotted as blue squares and red diamonds, respectively.  \emph{Right panel:} distribution of the target stars around NGC\,7492 using the same color code to identify the fields observed. The approximate field of view of VIMOS has been overplotted as reference.}
\label{seleccionplot}
     \end{center}
   \end{figure*}

NGC\,7492 is a poorly studied halo GC located at a heliocentric distance of $d_{\odot} = 26.2$\,kpc \citep[][]{Cote1991}, with Galactic coordinates ($\ell, b$) = (53.39$^{\circ}$,-63.48$^{\circ}$). Although its projected position and distance seem to be compatible with that of the LM10b model along the same line of sight, this cluster has a low  probability of being associated with the Sgr tidal stream, because on the predicted  differences in angular separation, heliocentric distance and radial velocity (LM10a). \cite{Carballo-Bello2014} unveiled an underlying system at the same heliocentric distance as of the cluster with wide-field photometry and suggested that at least a fraction of those stars belong to Sgr (see also Mu\~noz et al. 2017a). However, the stellar overdensities reported around NGC\,7492 \citep{Leon2000,Lee2004} and, more importantly, the tidal tails unveiled by \citet[][;hereafter N17]{Navarrete2017b} in Pan-STARRS\,1 survey data, may complicate the analysis of the underlying stellar system.  

In this paper, we derive kinematic information for a sample of stars in the surroundings of NGC\,7492 to assess whether this GC may have formed in Sgr and subsequently been accreted by the Milky Way.

\section{Observations and methodology} 
\label{photometry}

Our targets for spectroscopy have been selected from the photometric catalogs generated in a Megacam$@$CFHT and Megacam$@$Magellan survey of all outer Galactic halo satellites (Mu\~noz et al. 2017a). The color-magnitude diagram (CMD) for NGC\,7492 and those objects beyond 1.2 times the King tidal radius ($r_{\rm t}$) of the cluster are shown in \figref{seleccionplot}. We use a value of $r_{\rm t}$ as derived by \cite[][; $r_{\rm t}$ = 9.2\,arcmin]{Carballo-Bello2012}. As pointed out by \cite{Carballo-Bello2014}, a remarkable population of stars is found surrounding the cluster at large distances from its center with a main-sequence (MS) morphology similar to that of the GC, in the range $20 < g < 24$ and $0.2 < g-r < 0.9$. However, the small area covered aound the cluster in that study prevented them from reaching a conclusion about the nature of the underlying system. We selected targets in a box with $0.2 < g-r < 0.65$ and $16 < g < 22$ for stars with angular distances greater than 11\,arcmin from the center of NGC\,7492, including as well  turn-off (TO) stars of the underlying population and fore/background Milky Way objects. Two fields containing 71 stars were selected for targeting, with a total number of 31 and 40 stars observed in the fields 1 and 2, respectively (see \figref{seleccionplot}).   

Spectroscopic observations have been performed in service mode using the VIsible MultiObject Spectrograph (VIMOS) mounted at the 8.2\,m Very Large Telescope (Cerro Paranal, Chile) with the same setup used by \cite{Carballo-Bello2017}. The mid-resolution grism and the filter GG475 allowed a spectral coverage from 5000 to 8000\,\AA~ with a resolution $R = 580$.  Spectra are the result of a single exposure of 1740\,s with an average signal-to-noise ratio of 34, which have been extracted using the ESO \textsc{REFLEX} pipeline for VIMOS. We estimated the instrumental flexure by performing a second order correction of the wavelength calibration based on the position of several sky lines. The resulting offsets were removed from the individual spectra. We repeated this process iteratively until the shifts were negligible. 

Radial velocities\footnote{The derived radial velocities are included in an electronic table.} were derived by cross-correlating our normalized spectra with a list of templates  smoothed to the resolution of our results. The sample of templates used was compiled by \cite{Pickles1998} and we only considered MS or subgiant stars ranging from O to M spectral types in this procedure. The final radial velocity value was adopted from the template providing the best correlation coefficient. Radial velocities have been then corrected for the Earth motion relative to the heliocentric rest frame by using the \textsc{IRAF} task \textsc{rvcorrect} and their errors have been estimated from
the amplitude of the cross-correlation peak, using the prescriptions of
\cite{Tonry1979}. A mean error value of $<\sigma_{v_{\rm r}}> \sim 45$\,km\,s$^{-1}$ was found.

\section{Results and discussion}
\label{results}

\begin{figure}
     \begin{center}
      \includegraphics[scale=0.55]{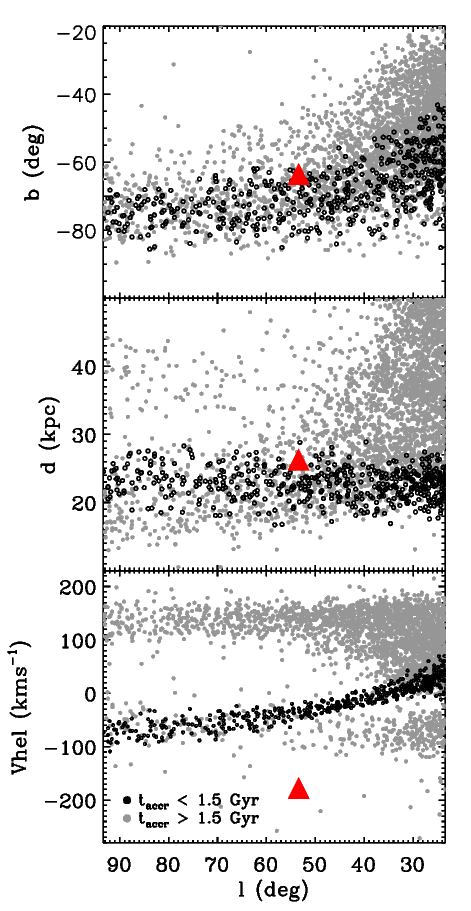}
      \caption[]{Projected path of the Sgr stream (up), heliocentric distance (middle) and radial velocity (bottom panel) as predicted by the LM10b model for $d_{\odot} < 50$\,kpc. The grey and black points are model particles corresponding to those stars accreted before (leading) and during the last (trailing) 1.5\,Gyr, respectively. The red triangle represents the position of NGC\,7492 on those planes.}
\label{modelo}
     \end{center}
   \end{figure}

We have visually matched the cluster CMD shown in \figref{seleccionplot} with a \cite{Dotter2008} isochrone with $t = 12$ and [Fe/H] = -1.8, as appropiate to NGC\,7492 \citep{Cote1991,Cohen2005,Forbes2010}. The adopted Galactic extinction values are $A_{\rm g} = 0.12$ and $A_{\rm r} = 0.08$, as derived from the \cite{Schlafly2011} extinction maps. The resulting heliocentric distance of NGC\,7492 is $d_{\odot} = 26.5 \pm 1.5$\,kpc, which is in good agreement with the estimate found for this cluster by \cite{Cote1991} and with the one derived by \cite{Figuera2013} within the errors. As for the hypothetical underlying system beyond $r=11$\,arcmin from the GC center, it presents a MS-TO magnitude similar to that of the cluster, which suggests that this stellar system and the cluster lie at the same heliocentric distance. Together with extra-tidal cluster members, the Sgr tidal stream is the main suspect of being responsible for the presence of those stars around NGC\,7492, as proposed by the numerical simulations obtained by LM10b and \citet[][hereafter P10]{Penarrubia2010}. In \figref{modelo} we show the projected position, heliocentric distance and expected radial velocity for that stream, according to the LM10b model for Sgr orbiting in a triaxial Galactic potential. Indeed, different sections of the stream seem to cross the area of the sky where NGC\,7492 is located. This would indicate that both cluster and tidal stream are spatially coincident. We fitted the CMD with the same isochrone used by \cite{Carballo-Bello2014} for Sgr ($t = 10$\,Gyr, ${\rm [Fe/H] } = -1.5$) and obtained a distance of $d_{\odot} \sim 26.8 \pm 1.7$\,kpc. We thus confirm that NGC\,7492 is immersed in the Sgr tidal stream.

In order to check whether some of our target stars could lie on
a tidal feature originating from NGC 7492, we generated two-dimensional density maps to search for tidal structures. We  performed a matched filter analysis following the description introduced by \cite{Rockosi2002} and widely used for the detection of extratidal features in Galactic GCs \citep[e.g.][N17]{Grillmair2006,Balbinot2011}. The central 1\,arcmin of NGC\,7492 were considered to derive the number density of stars in the CMD, while those stars beyond $r = 25$\,arcmin from the cluster center were used to sample the fore/background stellar populations. The CMD bin sizes used to compute those number densities were $\delta g=0.1$ and $\delta (g-r)=$0.05\,mag. 

\begin{figure}
     \begin{center}
      \includegraphics[scale=0.42]{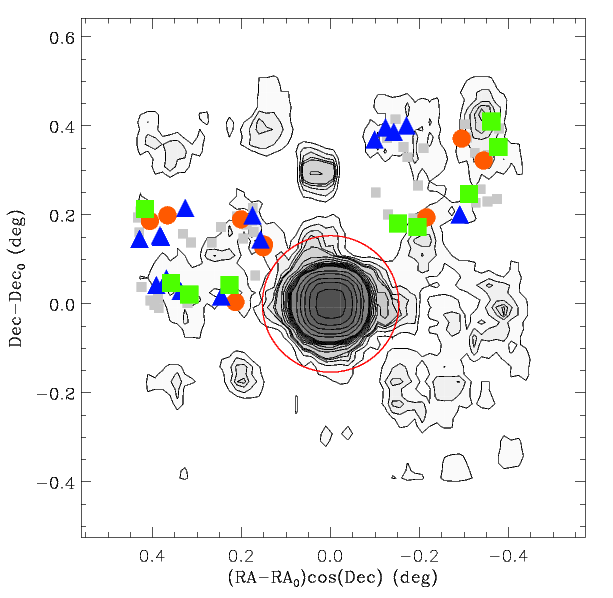}
      \caption[]{ Density map of NGC\,7492 stars as derived from the matched filter analysis, where North is up and East is to the left. The density contours displayed correspond to the number of standard deviations $\sigma$=[1, 1.5, 2, 3, 5, 10, 20, 30, 50, >50] above the mean background level. The target stars are overplotted as orange circles ($v_{\rm r} < -160$\,km\,s$^{-1}$), blue triangles ($-150 < v_{\rm r} {\rm \, [\,km\,s^{-1}]} < -90$),  green squares ($80 < v_{\rm r} {\rm \, [\,km\,s^{-1}]} < 150$) and small grey squares (rest of stars). The red line indicates the King tidal radius of NGC\,7492.}
\label{mapa}
     \end{center}
   \end{figure}

NGC\,7492 is a member of the group of ``tidally affected" GCs, according to the classification by \cite{Carballo-Bello2012}, so this cluster might be importantly distorted by its interaction with the Milky Way. We now focus on the density map generated using the procedure described above (see \figref{mapa}). Most of NGC\,7492's stars seem to be contained within the King tidal radius, which is almost fully filled by cluster members. Our density contours also show the presence of minor overdensities beyond $r_{\rm t}$, with a significance between 1 and 2$\sigma$ above the median background density. The orientation of these elongations, tentative tidal tails emerging from the cluster, is similar to that of the tails unveiled by N17 in Pan-STARRS1 survey data for NGC\,7492. Moreover, their density contours for NGC\,7492 show that this cluster would populate most of the Megacam field of view, with stars likely associated with the cluster found up to distances of 0.5\,deg from its center. Given our relatively small field of view, it is difficult to obtain a proper estimate of the contribution of fore/background populations to our density maps so we are not able to reproduce Navarrete's results with the same level of confidence. Even so, our NGC\,7492 isopleths may be used as a reference to establish the nature of the stars observed with VIMOS for this work.  

We have overplotted the position of the spectroscopic targets in the resulting density map shown in \figref{mapa}. Both groups of stars (fields\,1 and 2) are coincident in projected position with the  stellar overdensities revealed by the matched filter technique. In particular, field\,2 lies along the stellar arm unveiled by N17, so a contribution of cluster stars in our results is expected. However, even if stars belonging to NGC\,7492 are observed, the presence of the multiple wraps of Sgr predicted by the numerical simulations around this GC should be reflected in our radial velocity distribution. 

\begin{figure*}
     \begin{center}
      \includegraphics[scale=0.37]{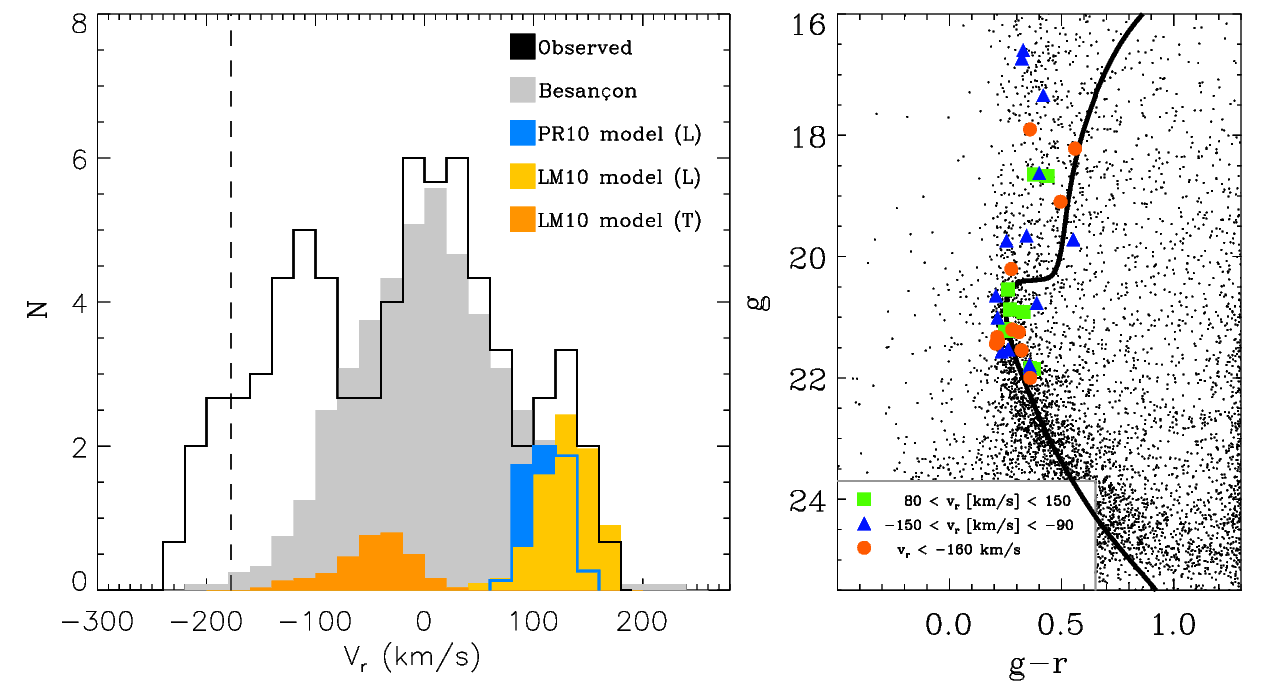}
      \caption[]{\emph{Left:} Radial velocity distribution obtained for the sample of stars around NGC\,7492 (black solid line). The blue area represents the velocities for Sgr stars according to P10, while the yellow and orange areas correspond to the leading and trailing arms of Sagitarius, respectively, as proposed by the LM10b model. The grey area corresponds to the expected velocity distribution of Milky Way stars along the same line of sight as NGC\,7492, as predicted by the Besan\c con model.  Grey, blue and orange areas have been arbitrarily scaled for better visualization. The dashed vertical line indicates the radial velocity of the GC measured by \cite{Cohen2005}. \emph{Right:} Megacam CMD corresponding to stars beyond 11\,arcmin from the center of NGC\,7492. The black solid line represents the isochrone for a stellar population with $t \sim 10$\,Gyr and [Fe/H] $\sim -1.5$ at the same heliocentric distance that NGC\,7492. Stars satisfying $v_{\rm r} < -160$, $-150 < v_{\rm r} < -90$ and $80 < v_{\rm r} < 150$\,km\,s$^{-1}$, are overplotted as orange circles, blue triangles and green squares, respectively. }
\label{histograma}
     \end{center}
   \end{figure*}

The histogram of velocities for the target stars was constructed using a bin size of 20\,km\,s$^{-1}$  and was then smoothed with a boxcar average with a width of 3 bins. The result is shown in \figref{histograma}. It is possible to distinguish at least 3 components in the heliocentric radial velocity distribution, with peaks with approximate mean velocities of $v_{\rm r} =$ -110, 0 and 125\,km\,s$^{-1}$, in descending order of number of stars. In order to identify the group of stars likely associated with the Milky Way stellar populations, we compare our results with the distribution obtained using the Besan\c con synthetic model \citep{Robin2003}. We performed 100 simulations using the default parametres for the position in the sky of NGC\,7492 and using a solid angle equivalent to 2\,deg$^{2}$. We randomly modified the velocities by adding values consistent with our mean observational error. The velocity distribution was derived by only including those synthetic stars in the same color-magnitud range as of the target stars and using the same bin size. The distribution predicted for Milky Way stars along this line of sight is arbitrarily scaled and overplotted in \figref{histograma} and confirms that the central group of stars around $v_{\rm r} \sim 0$\,km\,s$^{-1}$ is composed of target objects that are likely members of the Galactic disk and halo. 

As for the remaining kinematic components, we derive the distribution of velocities for particles in the P10 and LM10b models for the Sgr tidal stream in an area of 2\,deg~$\times$~2\,deg around the cluster position. The predicted signature of the stream along this line of sight is overplotted in \figref{histograma}. Both numerical simulations predict a concentration of stars around $v_{\rm r} \sim 110-130$\,km\,s$^{-1}$ belonging to the leading arm of the stream (according to the LM10b classification). This wrap of Sgr is predicted to be composed of stars  with accretion times $t_{\rm acc} > 1.5$\,Gyr and might be connected with the same wrap detected in the surroundings of Whiting\,1 by \cite{Carballo-Bello2017}. We thus confirm the detection of an old leading arm in the Sgr tidal stream, which is still detectable via multi-object spectroscopy over a wide field of view and located at $d_{\odot} \sim 26$\,kpc.     

The peak observed at $v_{\rm r} \sim -110$\,km\,s$^{-1}$ may be associated with the second component of Sgr predicted in this region of the sky. Following the nomenclature used by LM10b, a more recently accreted ($t_{\rm acc} < 0.5$\,Gyr) trailing arm section of the stream should be found around NGC\,7492 at a similar heliocentric distance but with $\langle{v_{\rm r}}\rangle \sim -50$\,km\,s$^{-1}$. Therefore, an important difference is found between our detection of the trailing arm of Sgr and the prediction made by the LM10b model. Given that no other halo substructure has been reported in this region, our results could be used to better constrain the orbital path (and kinematics) of the stream. The projected position in the sky for the stars found in both velocity peaks suggests that the stellar populations that we have identified as leading and trailing arms are found in both fields with no significant differences in their distribution (see \figref{mapa}). Although the model predicted more leading arm stars in this region of the sky, our histogram shows that the contribution of Sgr trailing arm stars is comparatively  more important in our sample. 

We also confirm that NGC\,7492's radial velocity \citep[$v_{\rm r} = -176.9$\,km\,s$^{-1}$;][]{Cohen2005} differs significantly from that of the Sgr tidal streams along this line of sight as previously predicted by LM10a. This indicates that NGC\,7492 is immersed in the Sgr tidal remnants but with a radial velocity that implies independent origins for both stellar systems. A few stars are found in our velocity distribution around the Cohen's velocity for the cluster, which was derived from the high-resolution spectra of 4 red-giant branch stars and is the measurement of reference in the literature for the kinematics of this globular. The position in the sky of the observed stars with $v_{\rm r} < -160$\,km\,s$^{-1}$ in the CMD is consistent with that of the isochrone corresponding to NGC\,7492 (see \figref{histograma}). Moreover, those stars seem to be concentrated in field 2, which is located along the stellar arm unveiled by N17. Therefore, our results  may confirm the findings of N17 about the presence of tidal tails in the field around this cluster. However, given the uncertainties of our velocities arising from the chosen instrumental setup, it is not possible to either unequivocally associate those stars with NGC\,7492 or estimate the number of cluster stars that are contributing the peak corresponding to the Sgr trailing arm.

\section{Conclusions}

We have derived radial velocities for a sample of 71 stars around the GC NGC\,7492 and distributed among 2 VIMOS fields. Our velocity distribution shows 3 peaks: Milky Way stars, Sgr leading and trailing arms members. These halo substructures seem to be located at the same heliocentric distance as that of NGC\,7492. 

The Sgr streams are found at $\langle{v_{\rm r}}\rangle \sim -110$ and 125\,km\,s$^{-1}$, with predicted accretion times $t_{\rm acc} < 0.5$ and $> 2$\,Gyr, respectively. According to the LM10b model, the latter represents a new detection of the old leading arm previously reported in a similar study with the same instrumental setup. As for the detection at negative velocities, it could  correspond to a section accreted during the last 0.5\,Gyr according to the numerical simulations available for the stream. Both components present a  kinematical signature different from that of the GC, which may support an independent origin for NGC\,7492, even when it is immersed in Sgr. A small number of stars  at $v_{\rm r} \sim -180$\,km\,s$^{-1}$ in our results seems to be associated with NGC 7492 members located along the tidal tail recently discovered, which is also revealed by our Megacam photometry.

\section*{Acknowledgements}
JAC-B acknowledges financial support from CONICYT-Chile FONDECYT Postdoctoral Fellowship \#3160502. JAC-B, MC and CN received support from the Ministry for the Economy, Development, and Tourism's Programa Iniciativa Cient\'ifica Milenio through grant IC120009, awarded to the Millennium Institute of Astrophysics (MAS) and from CONICYT's PCI program through grant DPI20140066. MC acknowledges additional support by Proyecto Basal PFB-06/2007 and FONDECYT grant \#1171273. CN acknowledges support from CONICYT-PCHA grant Doctorado Nacional 2015-21151643.   DMD acknowledges funding from Sonderforschungsbereich SFB 881 ``The Milky Way System" (subproject A2) of the German Research Foundation (DFG). R.R.M. acknowledges partial support from CONICYT Anillo project ACT-1122 and project BASAL PFB-$06$ as well as FONDECYT project N$^{\circ}1170364$. S.D. acknowledges support from Comit\'e Mixto ESO-GOBIERNO DE CHILE. MDM is supported by CONICYT, Programa de astronom\'ia, Fondo GEMINI, posici\'on Postdoctoral. Based on data products from observations made with ESO Telescopes at the La Silla Paranal Observatory under ESO programme ID 091.D-0446(D). Partially based on observations obtained at the CFHT, which is operated by the National Research Council of Canada, the Institut National des Sciences de l'\,Univers of the Centre National de la Recherche Scientifique of France, and the University of Hawaii.

\def\jnl@style{\it}                       
\def\mnref@jnl#1{{\jnl@style#1}}          
\def\aj{\mnref@jnl{AJ}}                   
\def\apj{\mnref@jnl{ApJ}}                 
\def\apjl{\mnref@jnl{ApJL}}               
\def\aap{\mnref@jnl{A\&A}}                
\def\mnras{\mnref@jnl{MNRAS}}             
\def\nat{\mnref@jnl{Nat.}}                
\def\iaucirc{\mnref@jnl{IAU~Circ.}}       
\def\atel{\mnref@jnl{ATel}}               
\def\iausymp{\mnref@jnl{IAU~Symp.}}       
\def\pasp{\mnref@jnl{PASP}}               
\def\araa{\mnref@jnl{ARA\&A}}             
\def\apjs{\mnref@jnl{ApJS}}               
\def\aapr{\mnref@jnl{A\&A Rev.}}          

\bibliographystyle{mn2e}
\bibliography{biblio}

\bsp	
\label{lastpage}
\end{document}